 \documentclass[
  aps,
  prd,
  twocolumn,
  showpacs, 
  showkeys,
  superscriptaddress,
  floatfix
  ]{revtex4}
 
 \usepackage[english]{babel}
  \usepackage[dvips]{color}
  \usepackage[dvips]{graphicx}
  \usepackage{latexsym}
  \usepackage{epsfig}
   \usepackage{amsmath}
  \usepackage{amssymb}
  \usepackage{multirow}
  \usepackage{tabularx}
  \usepackage{url}
  \usepackage{wrapfig}
  \usepackage{amsmath}
  \usepackage{amssymb}
  \usepackage{array}
  \usepackage{afterpage}
  \usepackage{url}
  \usepackage{multirow}
  \usepackage{tabularx}
  \usepackage{supertabular}
  \usepackage{longtable}
  \usepackage{rotating}
  \usepackage{natbib}  
  
  \newcommand{\MR}[2]{\multirow{#1}*{#2}}
  \newcommand{\MC}[3]{\multicolumn{#1}{#2}{#3}}

  \newcolumntype{C}{>{\centering}X}
  \newcolumntype{K}{>{\centering\arraybackslash\arraybackslash}X}

  \def\Frejus{Fr$\acute{\text{e}}$jus\,}
  \def\etal{\emph{et al}.}
  \def\deg{\text{o}}
  
\begin{document}

 \bibliographystyle{apsrev4-2}
  
 \title{High-energy spectra of the atmospheric neutrinos: predictions and measurements}         
 \author{A.~A.~Kochanov}
  \email{kochanov@iszf.irk.ru}
   \affiliation{Institute of Solar-Terrestrial Physics, \\ 
               Siberian Branch, Russian Academy of Sciences,
               RU-664033, Irkutsk, Russia}
  \affiliation{Irkutsk State University,
               RU-664003 Irkutsk, Russia}

   \author{A.~D.~Morozova}
 \email{anniemor@jinr.ru}
\affiliation{Irkutsk State University,
               RU-664003 Irkutsk, Russia}
    \affiliation{Dzhelepov Laboratory of Nuclear Problems, \\
               Joint Institute for Nuclear Research
               RU-141980 Dubna, Russia}

  \author{T.~S.~Sinegovskaya}
  \email{tanya@api.isu.ru}
  \affiliation{Irkutsk State Transport University,
               RU-664074, Irkutsk, Russia}

  \author{S.~I.~Sinegovsky}
  \email{sinegovsky@jinr.ru}
  \affiliation{Irkutsk State University,
               RU-664003 Irkutsk, Russia}
  \affiliation{Dzhelepov Laboratory of Nuclear Problems, \\
               Joint Institute for Nuclear Research,
               RU-141980 Dubna, Russia}
  \date{\today} 

  \begin{abstract}

Statistical analysis is performed of the atmospheric neutrino flux models as
compared with the data of Frejus, AMANDA-II, IceCube, ANTARES, and Super-Kamiokande experiments. 
The main objective is to characterize models of hadron-nucleus interactions from the point of view of the statistical significance of the atmospheric neutrino flux predictions compared to the measurements. 
The flux calculations were performed in the framework of the single computational scheme involving a set of  hadronic models combined with parameterizations of the primary cosmic rays spectrum by Hillas \& Gaisser and Zatsepin \& Sokolskaya. 
The analysis showed satisfactory agreement of the conventional $\nu_\mu$ flux calculations with the measurements.    
The prompt neutrinos conrtibution obtained with a set of charm production models (QGSM, SIBYLL 2.3c, PROSA,   GRRST, BEJKRSS, and GM-VFNS) is statistically negligible in the energy range covered by the neutrino telescopes.

  \end{abstract}



  \maketitle

  \section{Introduction}
 
 Decays of pions, kaons, and charmed particles produced in cosmic rays interactions with the Earth atmosphere  generate high-energy neutrinos which form an unavoidable background for detecting astrophysical neutrinos.     
 To date, the atmospheric muon and the electron neutrino spectra are measured in Frejus \cite{Daum:1994bf},  AMANDA-II \cite{Abbasi:2009nfa, Abbasi:2010qv}, 
  ANTARES \cite{Adrian-Martinez:2013bqq,Albert:2021pwz}, IceCube \cite{Abbasi:2010ie,Aartsen:2012uu, Aartsen:2014qna, Aartsen:2013eka,Aartsen:2015xup,Aartsen:2017nbu}, and Super-Kamiokande \cite{Richard:2015aua} experiments  %
 within the energy range $\sim 10$  GeV  through $600$ TeV.

 By the time, when  detector AMANDA at the South Pole was constructed,
  the Monte Carlo calculations of the atmospheric neutrinos (AN) spectra had been performed for the neutrino energies of no more than $10$ TeV \cite{Barr:2004br,Honda:2006qj,Honda:2011nf}.
  Later on, these results were used to reconstruct the events in the IceCube \cite{Aartsen:2013eka} and Super-Kamiokande experiments.

  The reference model for the AN spectra in IceCube  is based on the Monte Carlo calculation
  (for $E\leq 10$ TeV) \cite{Honda:2006qj}.  
 This model was extrapolated to energies beyond 10 TeV,  taking into consideration the knee of the primary cosmic ray spectrum with normalizing corrections depending on the energy \cite{Aartsen:2014qna}. 
 Thus, there is a necessity  of a consistent scheme for AN spectra calculations over a wide neutrino energy range. The scheme validation should be done via a thorough comparison with the experimental data.   

 Here we apply the $Z(E, h)$ method \cite{Naumov:2000au, Naumov:2001, Kochanov:2008pt, Sinegovskaya:2014pia}, developed to solve the high-energy atmospheric hadron cascade equations  and to compute the atmospheric muon and neutrino fluxes.
The  $Z(E, h)$ method enables us to compute atmospheric fluxes of hadrons, muons, and neutrinos for a non-power-law spectrum of cosmic rays, the non-scaling behavior of inclusive cross sections, and rising cross sections for inelastic hadron-nucleus collisions. 
The method have been tested \cite{Kochanov:2008pt, Sinegovsky:2010ijmpa} by comparing the calculated fluxes of high energy atmospheric hadrons  and  muons with the  data  of the past decade experiments.  The atmospheric muon spectra at various zenith angles have been thoroughly examined for wide energy range  \cite{Kochanov:2008pt, Kochanov:2013xea, Kochanov:2019jpcs}. 
The method enables one to estimate directly the effect of the primary cosmic ray spectrum and the hadronic interactions models on the absolute values of muon and neutrino fluxes without recourse to any normalizing factors.   

In this study, we analyze statistically the predicted atmospheric neutrinos spectra as compared  with measured ones, by using standard $\chi^2$ criterion.
This work continues and extends the topic touched upon in the conference talk~\cite{izvtas21}. \\


   

 \section{Spectra of atmospheric neutrinos} 
 
 The atmospheric neutrinos comprise two components, ``soft'' and ``hard'', clearly distinguishable in zenith-angle distributions and energy spectra. The anisotropic component originated from decays of pions and kaons has a softer  spectrum (``conventional'' neutrinos, CN). The quasi-isotropic flux of neutrinos  produced at higher energies, mainly in  decays of charmed hadrons   
($D^\pm$, $D^0$, $\overline{D}^0$, ${\rm\Lambda}^+_c$), features a harder spectrum
 (``prompt'' neutrinos, PN).
Due to a very short lifetime of charmed particles ($ \sim 10^{-12}-10^{-13} $ s), they decay close to their production point. This fact implies that the spectral index of the PN flux is a unit harder than the CN one.    %
 Calculations of the PN flux  are most uncertain due to wide spread predictions of the charm production models (see for example  \cite{Sinegovsky:2018vju}).
 Until now, the prompt component of the atmospheric neutrinos has not been identified in experiments.
A common expectation is that so-called ``crossing energy''  for prompt muon neutrinos is rather close to PeV. 

 \subsection{Conventional neutrinos} 

Conventional atmospheric neutrinos within $\sim 100$ GeV to $10$ PeV, produced in decays of $\pi^\pm$, $K^\pm$, $K^0_L$, and $K^0_S$ mesons were discussed in  \cite{Sinegovskaya:2014pia,Morozova:2017aic,Kochanov:2019owf,%
  Morozova:2019hbk,Morozova:2017eeo,Morozova:2017fof}.
The neutrino spectra were computed for a set of hadron-nucleus interaction models 
  QGSJET-II-03 \cite{Kalmykov:1997te,Ostapchenko:2004ss,Ostapchenko:2006wc},
  SIBYLL~2.1 \cite{Ahn:2009wx}, and model by Kimel \& Mokhov (KM) \cite{Kalinovsky:1989kk,Kimel:1974sn}, 
  which are used also in Monte Carlo simulations of extensive air showers produced by cosmic rays.
  We apply two  cosmic ray spectrum  models by Zatsepin and Sokolskaya (ZS) \cite{Zatsepin:2006ci} and 
 by Hillas and Gaisser \cite{Gaisser:2012zz} (chosen was the H3a version for mixed composition of the extragalactic component).
The ZS comprises contributions from three classes of Galaxy cosmic rays sources: isolated SNe exploding into  a random interstellar medium (ISM), high mass SNe exploding into a dense ISM 
(OB associations), and weak sources associated with novae explosions.
The ZS spectrum supported by direct measurements of ATIC-2 experiment \cite{Panov:2006kf,Panov:2011ak}  within 10 GeV--50 TeV serves, indeed, as an extrapolation of the CR spectrum beyond PeV (up to 100 PeV),
and thefore the calculated neutrino flux in case of ZS spectrum should be restricted to  $E_\nu < 10$ PeV.  

The Hillas and Gaisser  model \cite{Gaisser:2012zz} includes three classes of sources: supernovae remnants in the Galaxy,  Galaxy high-energy sources of unknown origin (that contribute to the cosmic ray flux between the knee (3 PeV) and the ankle (4 EeV)), and extragalactic astrophysical objects (Active Galactic Nuclei, sources of the gamma-ray bursts, and others).
The composite spectrum is formed of five nuclei groups (p, He, CNO, Mg-Si, and Mn-Fe). Each of the three
populations accelerates five nuclei groups, whose spectrum cuts off at a characteristic rigidity.

 Figures ~\ref{Fig-1}, \ref{Fig-2} present the coventional neutrinos fluxes, calculated with the hadronic models (QGSJET-II-03,  SIBYLL~2.1, and KM), together with the experimental data. 
 The main content of these figures is the comparison of calculated CN flux with  experiments. 
 However, the curves of the prompt neutrino spectra also shown in figures to exhibit distinctions between CN and PN fluxes.
Hereinafter,  $\nu_\mu$ and  $\nu_e$  designate the sums of neutrinos and antineutrinos, $\nu_\mu+\overline{\nu}_\mu$ and  $\nu_e+\overline{\nu}_e$, respectively.
 \subsection{Prompt neutrinos}

\subsubsection{Charm production models}

In this study, we use the  prompt neutrino (PN) contributions obtained with the charm production models:
  QGSM \cite{Sinegovsky:2018vju},  SIBYLL~2.3c \cite{Fedynitch:2018cbl}, 
  PROSA Collaboration \cite{Zenaiev:2019ktw, Garzelli:2017}, GRRST \cite{Gauld:2015kvh}, BEJKRSS \cite{Bhattacharya:2016jce}, and  GM-VFNS \cite{Benzke:2017yjn}.

 The non-perturbative quark-gluon string model (QGSM) was developed \cite{Kaidalov:1986zs, Kaidalov:2003au}  
to describe the soft and semihard hadronic processes at high energies. It has been applied to successfully describe the meson and baryon production in hadron-nucleon collisions.
The QGSM was one of the first models to estimate the atmospheric prompt  muon and  neutrino fluxes ~\cite{Bugaev:1989we,Bugaev:1998,NSS:1998}.
Here, we use the results of the prompt muon neutrino flux calculations at 1 TeV -- 100 PeV  performed  with the updated QGSM \cite{Sinegovsky:2018vju}. 
Prameters of the updated QGSM were examined by comparing the calculated cross sections for the charmed meson production  with the measurements in the LHCb and ALICE experiments. 
 Although the LHCb does not enable an unique choice of the QGSM parameters, the intercept of the Regge trajectory $\alpha_{\psi}(0)=-2.2$  appears a more preferable value versus  $\alpha_{\psi}(0)=0$. 
Another  QGSM free parameter is  the coefficient $a_{1}$ providing an unified description for the  $x\rightarrow0$ and $x\rightarrow1$ kinematic regions in the case of the leading fragmentation. 
There are no clear arguments for choosing $a_{1}$, and we calculate the PN flux for this parameter, varying from  $a_{1}=2$ and $a_{1}=30$ (shaded band in Figure~\ref{Fig-3}). 

 The SIBYLL~2.3c \cite{Fedynitch:2018cbl} was used to calculate the PN spectra  
 based on numerical solver for the system of the coupled cascade equations (Matrix Cascade Equations, $\rm{MCE_Q}$-method)  \cite{Fedynitch:2015, Fedynitch:2016}. 
The model for charm quark production is based on the LO QCD computations and the probability for the replacing $s$ quarks by $c$ ones in the fragmentation process.

The PROSA collaboration  \cite{Zenaiev:2019ktw, Garzelli:2017}
 has presented the thorough study of the atmospheric prompt neutrino problem with usage of the PROSA Monte Carlo event generator.  
The PROSA  computaions were based on the  LHCb and ALICE measurements of the charmed hadron production and improved constraints on the paron distribution functions (PDFs) in NLO QCD analysis using DIS and $pp$ collision data. 
 The prompt $\nu_\mu$ spectrum was obtained through the atmospheric cascade equations that describe the production and decay of secondary particles arising from cosmic ray interactions which produce finally the atmospheric muons and neutrinos.
  The  cascade equations admit approximate solutions in the $z$-moment approach with use of the superposition model for pA and AA interactions. 
The $z$-moments were calculated with the PROSA PDFs at the next-to-leading order (NLO) perturbative QCD (pQCD) in the fixed/variable flavour number schemes (FFNS/VFNS) consistent with the LHCb, ALICE and HERA measurements of charmed and beauty-flavoured hadrons.
It was shown that PDF uncertainties lead to the smaller flux uncertainty with respect to those 
arising from a choice of the QCD renormalization and factorization scales. 
The variations of phenomenological parameters of the charm fragmentation functions,
as well as the choice of the cosmic ray model (the CR compossition and spectrum) have also considerable impact on the flux uncertainty.

 The PN spectra predicted with GRRST model \cite{Gauld:2015kvh} are based on the Monte Carlo event generator using the $z$-moment approach to simulate of particles propagation and decays 
in the framework of NLO pQCD calculations through the same set of high energy charmed hadrons as in the PROSA approach. Cross sections of charmed particles production are obtained with the PDFs
integrated the measurements of charm production at the LHCb experiment.
Uncertainties of calculations are defined by the ``scale uncertainties'' of the NLO perturbative QCD and are reduced by NNLO calculations. Also the uncertainties of gluon PDF at small $x$ variable make a sizable contribution to the total errors of predicted PN, especially at $E_\nu > 1$ PeV. 

 The PN flux predicted with BEJKRSS model \cite{Bhattacharya:2016jce} was evaluated using the
 scheme for calculation of the charm production cross section, which comprises the NLO pQCD computations,
 the $k_T$ factorization approach with the low $x$ resummation, and the color dipole model comprising the gluon saturation. The QCD parameters were chosen to provide the best fit of the heavy quark production cross sections measured in RHIC and LHCb experiments.
The latest version of BEJKRSS model incorporates nuclear effects in the target PDFs, which usually are  neglected in the perturbative approach, although the nuclear shadowing effects may be considerable at very low $x$ region. It was found that the reduction of the neutrino  flux due to nuclear efects varies from $10\%$  to $50\%$ at the highest energies, depending on the used scheme. 
   
 The general-mass variable flavour number scheme (GM-VFNS) model \cite{Benzke:2017yjn} 
 bases on the NLO pQCD approach in which the matrix elements for the charm hadroproduction of light and heavy partons are combined with a set of fragmentation functions to describe the hadronization process (transition from partons to charmed hadrons). The scheme involves only one massive heavy quark $m_Q$, other quarks are massless.
  The GM-VFNS appropriates different approaches for calculation of cross sections: the FFNS scheme uses in low and intermediate $p_T$,  and zero mass VFNS framework does in high regions of $p_T$.
  The validity of the approach has been cross-checked by comparisons with data measured with the LHCb experiment.  Uncertainty of predicted PN spectra arises from  renormalization scale around central value
 (with a permanent level of half of order of PN magnitude)  and from the PDF uncertainty, which rapidly grows to two and a half orders of magnitude for very high energies.

 \subsubsection{PN flux predictions} 

 This work analysis concerns only atmospheric muon and electron neutrinos,  because tau neutrinos substantially are the prompt ones (originate from $D^\pm_s$ and $\tau$  decays) and they are suppressed by one order of magnitude \cite{Fedynitch:2018cbl}. 

\begin{figure} [!t]
  \centering
  \includegraphics[width=0.97\linewidth]
     {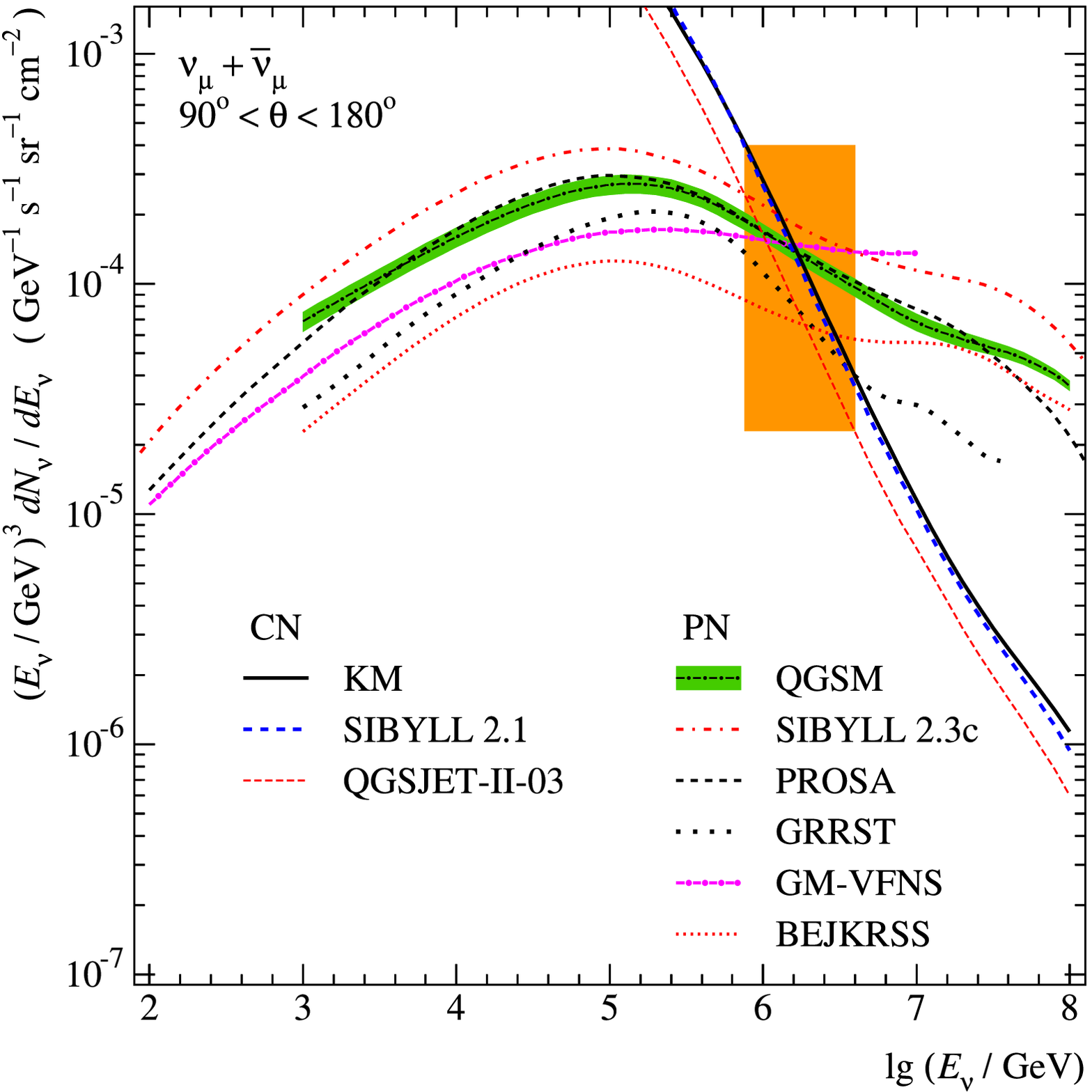} 
  \caption{Prompt $\nu_\mu$ spectra (scaled by factor $E^3_\nu$) at $\theta=0^\circ$  
   calculated with charm production models   
       QGSM \cite{Sinegovsky:2018vju}, SIBYLL~2.3c \cite{Fedynitch:2018cbl},
      GRRST \cite{Gauld:2015kvh}, BEJKRSS \cite{Bhattacharya:2016jce}, 
       GM-VFNS \cite{Benzke:2017yjn}, and PROSA \cite{Zenaiev:2019ktw}.
       The CN spectra averaged over zenith angles are also shown.  
The shaded rectangle displays the ``crossing energy'' range  predicted with different PN flux models.      
 All the fluxes are calculated for the H3a model of primary cosmic rays spectrum.    
          }
 \label{Fig-3}
  \end{figure}

The PN spectra are plotted in Figs.~\ref{Fig-3}, the calculations are performed  for the H3a spectrum of primary cosmic rays.  
Prompt electron neutrinos dominate over the conventional ones at energies beyond  $50$  TeV, while the prompt muon neutrinos become the dominant component at the PeV scale. 
 The PN energy spectra are calculated for the vertical direction (no averaging over zenith angles). The isotropic approximation provides a reasonable estimate at energies below 3 PeV where the PN flux weakly  depends on the zenith angle. The directions around the vertical are most suitable to reveal PN neutrinos because of the best PN/CN flux ratio.
 The green band shows the QGSM calculation uncertainty relating to varying free parameter $a_{1}$, that provides a unified description for the  kinematic regions $x\rightarrow 0$ and $x\rightarrow 1$ in the case, when the valence quarks participate in fragmentation \cite{Sinegovsky:2018vju}.
We do not show uncertainties of the spectra computed by the PROSA  \cite{Zenaiev:2019ktw} which absorb those of the rest models. All $\chi^2$ values are obtained for the central values of PN predictions.
  
 These models predictions display the spread of the prompt neutrino flux obtained for the H3a spectrum of cosmic rays, i.e. these models mark the approximate range of the PN contribution calculated with H3a spectrum (the use of the ZS spectrum weakly affects the range). 
 The PN fluxes  obtained with QGSM \cite{Sinegovsky:2018vju}, PROSA \cite{Zenaiev:2019ktw}  are very close to each other within a wide energy range (Fig.~\ref{Fig-3}).  The SIBYLL~2.3c \cite{Fedynitch:2018cbl} also predicts  the PN flux rather close to that of PROSA  and QGSM.
 Shaded area in Figure~\ref{Fig-3} shows the spread of model predictions for the median 
  ``crossing energy'', i.e.  the energy, above which the atmospheric neutrino flux is
    dominated by the PN component (see also Table~\ref{Tab:tab-4}). 
   All the fluxes are calculated for the H3a model of primary cosmic rays spectrum.  
     
  \begin{figure*}[!t]
  \centering
  \includegraphics[width=0.95\linewidth] {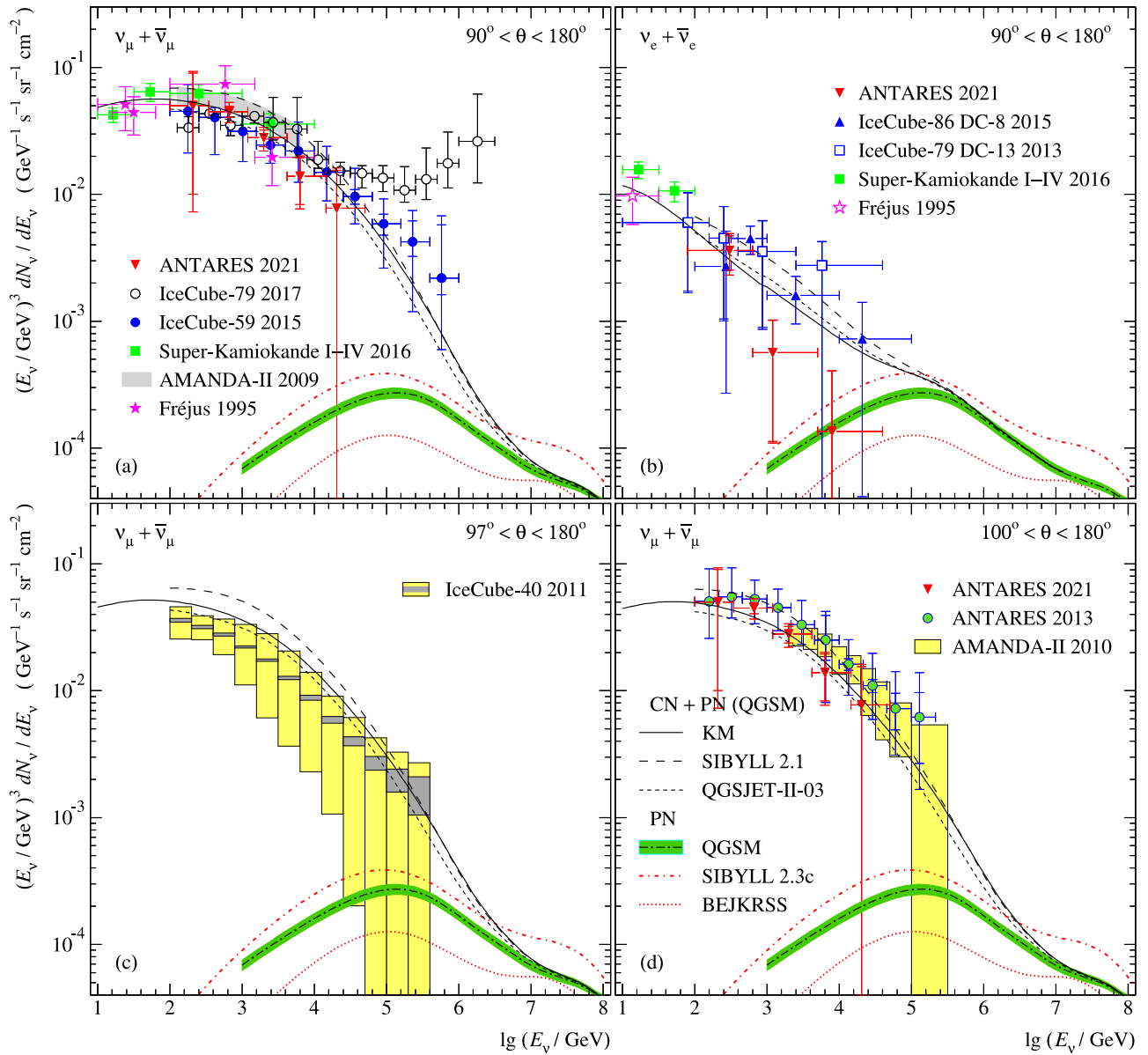}
   \caption{
  Calculated atmospheric neutrino spectra (scaled  by $E_\nu^3$) averaged over zenith angles, $\nu_\mu$ 
 (a), (c), (d), and  $\nu_e$ (b) compared to the  data of experiments: 
         \Frejus \cite{Daum:1994bf}, AMANDA-II \cite{Abbasi:2009nfa, Abbasi:2010qv}, 
           ANTARES \cite{Adrian-Martinez:2013bqq,Albert:2021pwz},
                      IceCube-40 \cite{Abbasi:2010ie},
           IceCube-59 \cite{Aartsen:2014qna}, IceCube-79 \cite{Aartsen:2012uu},
           IceCube-86 \cite{Aartsen:2015xup},
     and Super-Kamiokande I--IV \cite{Richard:2015aua}. 
  Error bars correspond to the uncertainties, including all statistical and systematic errors.
 Grey band in panel (a) indicates the 90\% CL from the forward-folding analysis
        of AMANDA-II 2009 \cite{Abbasi:2009nfa}.
  The total errors (yellow rectangles)  and statistical ones (dark rectangles) for the IceCube-40 data     and AMANDA-II 2010 \cite{Abbasi:2010qv} are shown in panels (c) and (d). 
  The spectra are calculated with H3a parameterization of the cosmic rays spectrum~\cite{Gaisser:2012zz}  for hadronic models KM (solid lines), SIBYLL~2.1 (dashed), and  QGSJET-II-03
   (short dashed).  Also shown are the PN spectra calculated with QGSM \cite{Sinegovsky:2018vju}, SIBYLL~2.3c \cite{Fedynitch:2018cbl}, and BEJKRSS \cite{Bhattacharya:2016jce}.
  }
  \label{Fig-1} 
  \end{figure*}
  \begin{figure*}[htb]
  \centering
  \includegraphics[width=0.95\linewidth]  %
  {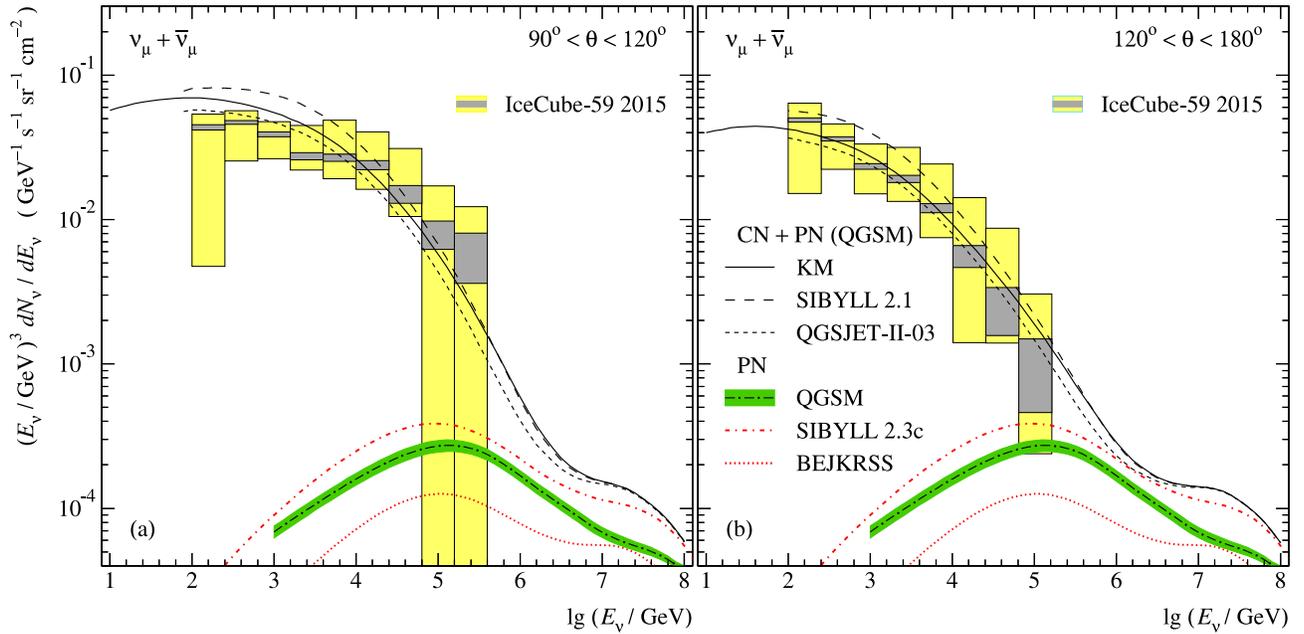}
   \caption{
   Calculated energy spectra of the atmospheric muon neutrinos  averaged over $90^\deg < \theta < 120^\deg$ (a),  and $120^\deg < \theta < 180^\deg$ (b) in comparison with the IceCube-59 data~\cite{Aartsen:2014qna}.  
  The total and statistical measurement errors are shown as light and dark shaded rectangles. 
  Notations of experimental errors and  theoretical predictions of the CN and PN fluxes
  are the same as in Figure~\ref{Fig-1}.
          }
  \label{Fig-2} 
  \end{figure*}

\section{Experimental data} 

The reconstructed neutrino energy spectra are derived from the neutrino telescopes with large statistical and systematical errors due to the restricted set of data and necessity to resort to sophisticated technique of handling the neutrino events. 
  The total errors of the  $\nu_\mu$ and $\nu_e$   spectra measured in the  \Frejus  \,   
  experiment \cite{Daum:1994bf} %
  at zenith angles  $90^\deg < \theta < 180^\deg$,  vary from $\sim$~26\% to $\sim$~55\% in low and high neutrino energies within the range of $0.25 < E_\nu \lesssim 10^3$ GeV.

  Preliminary results of the AMANDA-II experiment obtained at 90\% CL without zenith angles cuts were reported  in 2009 \cite{Abbasi:2009nfa} (grey band in the Figure \ref{Fig-1}\,(a)).
In the final AMANDA-II data \cite{Abbasi:2010qv}, muon tracks with $\theta < 100^\deg$
are removed from the analysis to minimize the atmospheric muon contamination of the neutrino sample.
The final sample contains 2972 neutrino-induced events.  
 The statistical errors of the $\nu_\mu$   spectra measured with the  AMANDA-II 
  (for 807 days between 2000 and 2003)  for $100^\deg < \theta < 180^\deg$
  are obtained of 18\% for low and 60\% for high neutrino energies.
  The systematical errors are 16\% within the entire energy range,
   $10^3 \lesssim E_\nu \lesssim 10^6$ GeV (yellow rectangles in 
   Figure \ref{Fig-1}\,(d)).

 
  The total errors of the neutrino spectra reconstructed in the  ANTARES 2013 \cite{Adrian-Martinez:2013bqq}
  (855 days in 2007--2011) for energies $10^2 < E_\nu \lesssim 10^6$ GeV
  are 30\% and 125\% for low and high energies, respectively.
 To suppress background events induced by atmospheric muons, zenith angles  were restricted to the range     $100^\deg < \theta < 180^\deg$.

The recent data published by the ANTARES Collaboration in 2021~\cite{Albert:2021pwz} were  collected in the period 2007--2017 in the energy range between $\sim 100$ GeV and  $ \sim 50$ TeV 
(5 bins for $\nu_\mu$ and 3 bins for  $\nu_e$); 
the zenith angles interval is $90^\deg < \theta < 180^\deg$.  The statistical uncertainties of the ANTARES 2021 reconstruction  are rather large:  $10\% - 100\%$ ($\nu_\mu$) and $30\% - 200\%$  ($\nu_e$). 
During of 3012 days of the livetime, about 130 $\nu_e$ and 850 $\nu_\mu$  events were reconstructed in the  instrumented volume of ANTARES detector.

The density of ANTARES optical modules is insufficient to reconstruct a considerable number of events induced by neutrinos at energies below 100 GeV.
The low statistics of the $\nu_e$ events prevents from testing charm production models above tens of TeV, i.e. in the energy range, where one could expect an appreciable contribution of the prompt electron neutrinos. 
The energy estimate for the semi-contained events relative to the through-going ones reduces the overall uncertainty of the measured flux as compared  with  ANTARES 2013 measurements. 
The ANTARES 2021 $\nu_\mu$ flux  is close to that of IceCube-40 (2011) and IceCube-59 (2015), and $20\% - 25\%$ below the flux reported in the ANTARES 2013 measurement.

  The total experimental errors of the neutrino spectra reconstructed
  in the Super-Kamiokande \cite{Richard:2015aua} 
  (operated intermittently since  1996, but the data sets used in this paper include the data until 2015)
  vary from 15\% to 19\% ($\nu_e$) and  $15\% - 21\%$ ($\nu_\mu$) 
   in the energy range $0.25 < E_\nu < 10^4$ GeV.
 
  In the period $2008-2009$, the IceCube detector was operated in the 40-strings configuration \cite{Abbasi:2010ie}. 
  The results of the measurements were presented for
  three zenith angles intervals:  $97^\deg < \theta < 124^\deg$,
  $124^\deg < \theta < 180^\deg$, and   $97^\deg < \theta < 180^\deg$ (joint interval).
  The authors of the experiment have published the errors only  for the joint interval.
  The total errors for the joint analysis were estimated  from 21\% to 158\% for low and high energies
  in the range $10^2 < E_\nu \lesssim 10^6$ GeV.

%
 The first data on the atmospheric $\nu_e$  flux in TeV energy range were obtained  in  $2010-2011$ using the DeepCore infill array in the IceCube-79 (the 79-strings configuration) \cite{Aartsen:2012uu}.  DeepCore included the six specialized strings and the seven adjacent standard IceCube strings (DeepCore-13) and allowed reducing the energy threshold to $\sim 10$ GeV. 
  Four data points of the reconstructed atmospheric $\nu_e$ spectrum were obtained from a selected data sample of $496 \pm 66 $ (stat) $\pm 88 $ (syst) cascade  events observed in 281 days of data. They included $\nu_e$ charge currrent interactions and neutral current interactions  of neutrinos of all flavors. 
 The experimental errors of measured $\nu_e$  spectra are from 54\% to 100\% in the energy range 
 80 GeV  -- 6 TeV.
  %
 
  In  $2011-2012$, a new IceCube analysis of the $\nu_e$ spectrum was based on the data taken 
  for $97^\deg < \theta < 180^\deg$  with the full 86-string configuration during 332.3 days of livetime
 \cite{Aartsen:2015xup}.
   The total errors of data are estimated as  $25\%-94\%$ for energies $10^2 < E_\nu \lesssim 10^5$ GeV.  
 Whereas the information on zenith angles cuts is not entirely clear, we do not apply any cuts in our analysis.
  The IceCube-59 experimental data for $\nu_\mu$ were taken in $2009-2010$ with the 59-string 
  configuration of the detector \cite{Aartsen:2014qna}. 
  Like in the IceCube-40,  the analysis of events addresses three intervals of zenith angles,
  $90^\deg < \theta < 120^\deg$,  $120^\deg < \theta < 180^\deg$, and  $90^\deg < \theta < 180^\deg$.
  The total errors are estimated from $\sim$~25\% to $\sim$~250\%
  for energies $10^2 < E_\nu \lesssim 10^6$ GeV.
  We do not use the IceCube-79 data \cite{Aartsen:2017nbu} in our analysis, 
  because they contain uncertain admixture of astrophysical neutrinos. 

  Thus, the total data set for the statistical analysis for
  $\nu_\mu$ spectra contains 54 data points measured with 
  \Frejus (4 points), AMANDA-II (9 points), IceCube-40 (12 points), ANTARES 2013 (10 points), 
   ANTARES 2021 (5 points),   Super-Kamiokande (4 points), and IceCube-59 (10 points) 
   at zenith angles in the interval  $90^\deg < \theta < 180^\deg$.
The set for the analysis of $\nu_e$ spectra includes 11 data points obtained in  
 Super-Kamiokande \cite{Richard:2015aua}, IceCube-79 \cite{Aartsen:2012uu},
 IceCube-86, and  ANTARES 2021 \cite{Albert:2021pwz}.
Figures \ref{Fig-1} and \ref{Fig-2} show the measured and predicted spectra calculated with
  the models tested in our analysis.
  Figure~\ref{Fig-1}(b) shows the experimental data for $\nu_e$ spectra measured 
  in \Frejus 1995 \cite{Daum:1994bf},
  Super-Kamiokande \cite{Richard:2015aua} 
  for zenith angles $90^\deg < \theta < 180^\deg$
   along with the predicted spectra  averaged over the narrowed angle interval, $97^\deg < \theta < 180^\deg$.


 \section {Statistical analysis}  


 A set of $n$ independent measurements $\Phi_i$ (energy spectrum) at points $E_i$ (energy) is considered as Gaussian distributed with the mean $\mu(E_i;\vec{\alpha},\vec{\beta})$ and known variance $\sigma^2_i$.   The goal of the statistical analysis is to construct estimators for the unknown parameters $\vec{\alpha},\vec{\beta}$. In our case, $\vec{\alpha}$ stands for hadronic models ($\alpha_j, j=1, 2, 3$)  and  $\vec{\beta}$ labels models of cosmic ray spectrum  ($\beta_k, k=1, 2$).  That is, index $j$ implies a hadronic interaction model KM, QGSJET-II-03 and SIBYLL 2.1, and $j$ marks CR models H3a and ZS.   

For the statistical analysis \cite{pval1, pval2}  we use $\chi^2$ values 
 \begin{equation} \chi^2 (\vec{\alpha}, \vec{\beta})=
 \sum_{i=1}^{\rm ndf}\frac{\left(\Phi^{\rm exp}_i-<\phi^{(\vec{\alpha},\,  \vec{\beta})}_i>\right)^2}{(\delta\Phi^{\rm exp}_i)^2},
\end{equation}  
 where $\Phi^{\rm exp }_i$ is the detected neutrino flux for $i$-th energy bin;
  $\phi^{(\vec{\alpha}, \vec{\beta})}_i$ is the calculated one for the chosen flux model ($j,k$); ndf is the  number of the data bins (data points) index $i$ enumerates the measured mean values $E_i$; 
  $\delta\Phi^{\rm exp}_i$ designates experimental errors (considering the systematic and statistical uncertainties).
For each bin $\Phi(E_\nu)\equiv <dN_\nu/dE_\nu>_\theta $ denotes the differential neutrino flux (the energy spectrum) averaged  over zenith angles.
 The predicted neutrino flux was averaged over energy in the  $i$-th  bin: 
  \begin{equation}
	<\phi^{(\vec{\alpha},\,  \vec{\beta})}(\bar{E_i})>=\frac{1}{\Delta E_i}\int^{E_{i+1}}_{E_{i}}\phi^{(\vec{\alpha},\,  \vec{\beta})}(E)dE.
\end{equation}
%
\begin{table*}[htb]
  \caption{
   $\chi^2/{\rm ndf}$ values for the predicted  $\nu_\mu$ spectra versus  the experimental data. 
   Calculations are made for the H3a and ZS (in brackets) cosmic ray spectrum.}
 \label{Tab:tab-1}   
  \center{
  \begin{tabularx}{\linewidth}{@{\qquad}l@{\quad}|C|CCK}                                                          \hline\hline\noalign{\smallskip}
  Experiment,          & $\chi^2$ (CN) &       & $\chi^2$ (CN+PN)    &          \\
  CN $|$ PN   models   &               & QGSM  & SIBYLL~2.3c         & BEJKRSS         \\
  \noalign{\smallskip}\hline      \noalign{\smallskip}
    \MC{5}{l}{{\bf \Frejus 1995}
             \cite{Daum:1994bf},
             $ E_\nu \lesssim 10^3$ GeV,
             $90^\deg < \theta < 180^\deg$}                                           \\
         KM           & \MC{4}{c}{2.30/ 4 = 0.57 (2.77/ 4 = 0.69)}                    \\
         QGSJET-II-03 & \MC{4}{c}{0.23/ 2 = 0.11 (0.36/ 2 = 0.18)}                    \\
         SIBYLL~2.1   & \MC{4}{c}{6.78/ 2 = 3.39 (6.77/ 2 = 3.39)}                    \\        
  \\
  \MC{5}{l}{{\bf AMANDA-II 2010}
            \cite{Abbasi:2010qv},
            $10^3 \lesssim E_\nu \lesssim 10^6$ GeV,
            $100^\deg < \theta < 180^\deg$}                                                \\
  \MR{1}{KM}          & 21.4/ 9 = 2.38 & 20.5/ 9 = 2.28 & 20.1/ 9 = 2.23 & 21.0/ 9 = 2.33  \\
                      &(29.1/ 9 = 3.23)&(28.1/ 9 = 3.12)&(27.6/ 9 = 3.07)&(28.6/ 9 = 3.18) \\
  \MR{1}{QGSJET-II-03}& 31.4/ 9 = 3.49 & 30.3/ 9 = 3.36 & 29.7/ 9 = 3.30 & 30.9/ 9 = 3.43  \\
                      &(39.5/ 9 = 4.39)&(38.3/ 9 = 4.26)&(37.8/ 9 = 4.20)&(39.0/ 9 = 4.33) \\
  \MR{1}{SIBYLL~2.1}  & 6.52/ 9 = 0.72 & 5.94/ 9 = 0.66 & 5.70/ 9 = 0.63 & 6.25/ 9 = 0.69  \\
                      &(11.5/ 9 = 1.28)&(10.8/ 9 = 1.20)&(10.4/ 9 = 1.16)&(11.1/ 9 = 1.23) \\
 \\
  \MC{5}{l}{{\bf IceCube-40 2011}
            \cite{Abbasi:2010ie},
            $10^2 < E_\nu \lesssim 10^6$ GeV,
            $97^\deg < \theta < 180^\deg$}                                                 \\
  \MR{1}{KM}          & 0.78/12 = 0.06 & 0.87/12 = 0.07 & 0.92/12 = 0.08 & 0.82/12 = 0.07  \\
                      &(0.70/12 = 0.06)&(0.72/12 = 0.06)&(0.75/12 = 0.06)&(0.71/12 = 0.06) \\
  \MR{1}{QGSJET-II-03}& 0.64/12 = 0.05 & 0.66/12 = 0.06 & 0.66/12 = 0.05 & 0.65/12 = 0.05  \\
                      &(0.34/12 = 0.03)&(0.32/12 = 0.03)&(0.30/12 = 0.03)&(0.33/12 = 0.03) \\
  \MR{1}{SIBYLL~2.1}  & 13.3/12 = 1.10 & 13.5/12 = 1.13 & 13.7/12 = 1.14 & 13.4/12 = 1.12  \\
                      &(13.4/12 = 1.12)&(13.5/12 = 1.13)&(13.7/12 = 1.14)&(13.5/12 = 1.13) \\
  \\
  \MC{5}{l}{{\bf ANTARES 2013}
            \cite{Adrian-Martinez:2013bqq},
            $10^2 < E_\nu \lesssim 10^6$ GeV,
            $100^\deg < \theta < 180^\deg$}                                                \\
  \MR{1}{KM}          & 4.46/10 = 0.45 & 4.23/10 = 0.42 & 4.12/10 = 0.41 & 4.35/10 = 0.44  \\
                      &(5.35/10 = 0.53)&(5.10/10 = 0.51)&(4.98/10 = 0.50)&(5.24/10 = 0.52) \\
  \MR{1}{QGSJET-II-03}& 7.17/10 = 0.72 & 6.90/10 = 0.69 & 6.77/10 = 0.68 & 7.05/10 = 0.70  \\
                      &(8.19/10 = 0.82)&(7.91/10 = 0.79)&(7.77/10 = 0.78)&(8.06/10 = 0.81) \\
  \MR{1}{SIBYLL~2.1}  & 1.58/10 = 0.16 & 1.43/10 = 0.14 & 1.37/10 = 0.14 & 1.51/10 = 0.15  \\
                      &(2.35/10 = 0.23)&(2.18/10 = 0.22)&(2.10/10 = 0.21)&(2.27/10 = 0.23) \\ 
 \\
   \MC{5}{l}{{\bf Super-Kamiokande I--IV 2016}
             \cite{Richard:2015aua},
             $E_\nu < 10^4$ GeV,
             $90^\deg < \theta < 180^\deg$}                                                \\
     KM            & \MC{4}{c}{3.65/ 4 = 0.91 (4.29/ 4 = 1.07)}                    \\
     QGSJET-II-03 & \MC{4}{c}{4.01/ 2 = 2.01 (4.00/ 2 = 2.00)}                    \\  
     SIBYLL~2.1   & \MC{4}{c}{1.38/ 2 = 0.69 (1.26/ 2 = 0.63)}                    \\ 
                   
                       \noalign{\smallskip}       \hline\noalign{\smallskip}
 \\
  \MC{5}{l}{{\bf Combined data}
            \cite{Daum:1994bf,             
                  Abbasi:2010qv,           
                  Adrian-Martinez:2013bqq, 
                  Richard:2015aua,         
                  Abbasi:2010ie}           
            $ E_\nu \lesssim 10^6$ GeV}                                              \\
  \MR{1}{KM}          & 32.6/39 = 0.84 & 31.5/39 = 0.81 & 31.1/39 = 0.80 & 32.1/39 = 0.82  \\
                      &(42.2/39 = 1.08)&(41.0/39 = 1.05)&(40.4/39 = 1.04)&(41.6/39 = 1.07) \\
  \MR{1}{QGSJET-II-03}& 43.4/35 = 1.24 & 42.1/35 = 1.20 & 41.3/35 = 1.18 & 42.8/35 = 1.22  \\
                      &(52.4/35 = 1.50)&(50.9/35 = 1.45)&(50.2/35 = 1.43)&(51.7/35 = 1.48) \\
  \MR{1}{SIBYLL~2.1}  & 29.6/35 = 0.84 & 29.1/35 = 0.83 & 29.0/35 = 0.83 & 29.3/35 = 0.84  \\
                      &(35.3/35 = 1.01)&(34.6/35 = 0.99)&(34.4/35 = 0.98)&(35.0/35 = 1.00) \\ \noalign{\smallskip}
  \noalign{\smallskip}\hline\hline\noalign{\smallskip}
   \end{tabularx}}
  \end{table*}

 \begin{table*}[htb!]
  \caption{
          $\chi^2/{\rm ndf}$ values obtained for predicted $\nu_\mu$ spectra and 
          measured ones in the IceCube and ANTARES experiments.}
 %
  \label{Tab:tab-2}     
    \center{
  \begin{tabularx}{\linewidth}{@{\qquad}l@{\quad}|C|CCK}                                                          \hline\hline\noalign{\smallskip}
%
  Experiment, & $\chi^2$ (CN) &                 & $\chi^2$ (CN+PN)    &          \\
  CN $|$ PN models   &              & QGSM  & SIBYLL~2.3c        & BEJKRSS         \\
  \noalign{\smallskip}\hline      \noalign{\smallskip}
 
  \MC{5}{l}{{\bf IceCube-59 2015}
         \cite{Aartsen:2014qna},           
            $10^2 < E_\nu \lesssim 10^6$ GeV,
            $90^\deg < \theta < 120^\deg$}                                                 \\
  \MR{2}{KM}          & 11.0/ 9 = 1.22 & 10.9/ 9 = 1.22 & 11.0/ 9 = 1.22 & 10.9/ 9 = 1.21  \\
                      &(12.7/ 9 = 1.41)&(12.6/ 9 = 1.40)&(12.7/ 9 = 1.41)&(12.6/ 9 = 1.40) \\
  \MR{2}{QGSJET-II-03}& 4.60/ 9 = 0.51 & 4.52/ 9 = 0.50 & 4.51/ 9 = 0.50 & 4.56/ 9 = 0.51  \\
                      &(4.89/ 9 = 0.53)&(4.70/ 9 = 0.52)&(4.69/ 9 = 0.52)&(4.75/ 9 = 0.53) \\
  \MR{2}{SIBYLL~2.1}  & 35.1/ 9 = 3.90 & 35.1/ 9 = 3.90 & 35.2/ 9 = 3.91 & 35.1/ 9 = 3.90  \\
                      &(36.2/ 9 = 4.02)&(36.2/ 9 = 4.02)&(36.3/ 9 = 4.03)&(36.2/ 9 = 4.02) \\
 \\
  \MC{5}{l}{{\bf IceCube-59 2015}
             \cite{Aartsen:2014qna},
             $10^2 < E_\nu \lesssim 10^6$ GeV,
             $120^\deg < \theta < 180^\deg$}                                               \\
  \MR{2}{KM}          & 0.97/ 8 = 0.12 & 1.26/ 8 = 0.16 & 1.43/ 8 = 0.18 & 1.10/ 8 = 0.14  \\
                      &(1.22/ 8 = 0.15)&(1.50/ 8 = 0.19)&(1.66/ 8 = 0.21)&(1.34/ 8 = 0.17) \\
  \MR{2}{QGSJET-II-03}& 0.41/ 8 = 0.05 & 0.55/ 8 = 0.07 & 0.64/ 8 = 0.08 & 0.47/ 8 = 0.06  \\
                      &(0.52/ 8 = 0.07)&(0.67/ 8 = 0.08)&(0.76/ 8 = 0.10)&(0.58/ 8 = 0.07) \\
  \MR{2}{SIBYLL~2.1}  & 10.6/ 8 = 1.33 & 11.1/ 8 = 1.39 & 11.4/ 8 = 1.42 & 10.8/ 8 = 1.35  \\
                      &(11.5/ 8 = 1.44)&(12.0/ 8 = 1.50)&(12.2/ 8 = 1.53)&(11.7/ 8 = 1.46) \\
  \\
  \MC{5}{l}{{\bf IceCube-59 2015}
             \cite{Aartsen:2014qna},
             $10^2 < E_\nu \lesssim 10^6$ GeV,
             $90^\deg < \theta < 180^\deg$}                                                \\
  \MR{2}{KM}          & 4.79/10 = 0.48 & 4.49/10 = 0.45 & 4.40/10 = 0.44 & 4.64/10 = 0.46  \\
                      &(5.82/10 = 0.58)&(5.51/10 = 0.55)&(5.42/10 = 0.54)&(5.67/10 = 0.57) \\
  \MR{2}{QGSJET-II-03}& 3.58/10 = 0.36 & 3.15/10 = 0.32 & 3.00/10 = 0.30 & 3.38/10 = 0.34  \\
                      &(3.89/10 = 0.39)&(3.46/10 = 0.35)&(3.31/10 = 0.33)&(3.69/10 = 0.37) \\
  \MR{2}{SIBYLL~2.1}  & 18.0/10 = 1.80 & 17.8/10 = 1.78 & 17.8/10 = 1.78 & 17.9/10 = 1.79  \\
                      &(18.8/10 = 1.88)&(18.6/10 = 1.86)&(18.6/10 = 1.86)&(18.7/10 = 1.87) \\ \noalign{\smallskip}       \hline\noalign{\smallskip}
  \MC{5}{l}{{\bf ANTARES 2021}
             \cite{Albert:2021pwz},  
             $10^2 < E_\nu \lesssim 5 \times 10^4$ GeV,
             $90^\deg < \theta < 180^\deg$}\\
  \MR{2}{KM}          & 1.74/ 5 = 0.35 & 1.80/ 5 = 0.36 & 1.84/ 5 = 0.37 & 1.76/ 5 = 0.35  \\
                      &(1.72/ 5 = 0.34)&(1.76/ 5 = 0.35)&(1.79/ 5 = 0.36)&(1.74/ 5 = 0.35) \\
  \MR{2}{QGSJET-II-03}& 0.19/ 5 = 0.04 & 0.20/ 5 = 0.04 & 0.20/ 5 = 0.04 & 0.19/ 5 = 0.04  \\
                      &(0.12/ 5 = 0.02)&(0.11/ 5 = 0.02)&(0.10/ 5 = 0.02)&(0.12/ 5 = 0.02) \\
  \MR{2}{SIBYLL~2.1}  & 20.0/ 5 = 4.00 & 20.2/ 5 = 4.04 & 20.4/ 5 = 4.08 & 20.1/ 5 = 4.02  \\
                      &(19.4/ 5 = 3.88)&(19.6/ 5 = 3.91)&(19.7/ 5 = 3.94)& 19.5/ 5 = 3.90  \\ \noalign{\smallskip}
 \\
  \MC{5}{l}{{\bf ANTARES 2013} \cite{Adrian-Martinez:2013bqq} \&
            {\bf ANTARES 2021} \cite{Albert:2021pwz}
             $10^2 < E_\nu \lesssim 10^6$ GeV} \\
  \MR{2}{KM}          & 6.20/15 = 0.41 & 6.03/15 = 0.40 & 5.96/15 = 0.40 & 6.11/15 = 0.41  \\
                      &(7.07/15 = 0.47)&(6.86/15 = 0.46)&(6.77/15 = 0.45)&(6.98/15 = 0.47) \\
  \MR{2}{QGSJET-II-03}& 7.37/15 = 0.49 & 8.66/15 = 0.58 & 6.97/15 = 0.46 & 7.24/15 = 0.48  \\
                      &(8.31/15 = 0.55)&(8.02/15 = 0.53)&(7.87/15 = 0.52)&(8.18/15 = 0.55) \\
  \MR{2}{SIBYLL~2.1}  & 21.6/15 = 1.44 & 21.6/15 = 1.44 & 21.8/15 = 1.45 & 21.6/15 = 1.44  \\
                      &(21.8/15 = 1.45)&(21.8/15 = 1.45)&(21.8/15 = 1.45)&(21.8/15 = 1.45) \\
                   
\noalign{\smallskip}       \hline\noalign{\smallskip}
\\
  \MC{5}{l}{{\bf Combined data}
            \cite{Daum:1994bf,             
                  Abbasi:2010qv,           
                  Adrian-Martinez:2013bqq, 
                  Richard:2015aua,         
                  Abbasi:2010ie,           
                  Aartsen:2014qna,         
                  Albert:2021pwz}          
            $ E_\nu \lesssim 10^6$ GeV}                                              \\
  \MR{2}{KM}          & 39.1/54 = 0.72 & 37.8/54 = 0.70 & 37.3/54 = 0.69 & 38.5/54 = 0.71  \\
                      &(49.8/54 = 0.92)&(48.2/54 = 0.89)&(47.6/54 = 0.88)&(49.0/54 = 0.91) \\
  \MR{2}{QGSJET-II-03}& 47.2/50 = 0.94 & 47.0/50 = 0.94 & 44.5/50 = 0.89 & 46.4/50 = 0.93  \\
                      &(56.4/50 = 1.13)&(54.4/50 = 1.09)&(53.6/50 = 1.07)&(55.5/50 = 1.11) \\
  \MR{2}{SIBYLL~2.1}  & 67.6/50 = 1.35 & 67.1/50 = 1.34 & 67.2/50 = 1.34 & 67.3/50 = 1.35  \\
                      &(73.5/50 = 1.47)&(72.8/50 = 1.46)&(72.7/50 = 1.45)&(73.2/50 = 1.46) \\ \noalign{\smallskip} 
                       \noalign{\smallskip}\hline\hline\noalign{\smallskip}
  \end{tabularx}}
  \end{table*}

  \begin{table*}[htb!]
  \caption{
        $\chi^2/{\rm ndf}$ values calculated for the measured and predicted
            $\nu_e$ spectra. 
          }     
   \label{Tab:tab-3}     
  \center{
  \begin{tabularx}{\linewidth}{@{\qquad}l@{\quad}|C|CCK}                                                          \hline\hline\noalign{\smallskip}
  
  Experiment, & $\chi^2$ (CN) &                 & $\chi^2$ (CN+PN)    &          \\
  CN $|$ PN models  &              & QGSM  & SIBYLL~2.3c        & BEJKRSS         \\
  \noalign{\smallskip}\hline      \noalign{\smallskip}
  \MC{5}{l}{{\bf Super-Kamiokande I--IV 2016}  \cite{Richard:2015aua},           
  
            $E_\nu < 100$ GeV,
            $90^\deg < \theta < 180^\deg$}                                                 \\
         KM           & \MC{4}{c}{ 8.44/ 2 = 4.22 (7.21/ 2 = 3.61)}                        \\
 \\
  \MC{5}{l}{{\bf IceCube-79 DeepCore-13 2013}  \cite{Aartsen:2012uu},
            $10 < E_\nu \lesssim 10^5$ GeV,  
            $90^\deg < \theta < 180^\deg$}                                                 \\
  \MR{2}{KM}          & 22.5/ 4 = 5.63 & 22.4/ 4 = 5.60 & 22.4/ 4 = 5.60 & 22.5/ 4 = 5.63  \\
                      &(24.9/ 4 = 6.23)&(24.8/ 4 = 6.20)&(24.8/ 4 = 6.20)&(24.9/ 4 = 6.23) \\
  \MR{2}{QGSJET-II-03}& 1.11/ 3 = 0.37 & 1.01/ 3 = 0.34 & 0.95/ 3 = 0.32 & 1.07/ 3 = 0.36  \\
                      &(1.18/ 3 = 0.39)&(1.08/ 3 = 0.36)&(1.01/ 3 = 0.34)&(1.14/ 3 = 0.38) \\
  \MR{2}{SIBYLL~2.1}  & 0.68/ 3 = 0.23 & 0.60/ 3 = 0.20 & 0.57/ 3 = 0.19 & 0.65/ 3 = 0.22  \\
                      &(0.82/ 3 = 0.27)&(0.74/ 3 = 0.25)&(0.71/ 3 = 0.24)&(0.78/ 3 = 0.26) \\ \noalign{\smallskip}
  \\
  \MC{5}{l}{{\bf IceCube-86 DeepCore-8 2015}   \cite{Aartsen:2015xup},
            $10^2 < E_\nu \lesssim 10^5$ GeV,
            $90^\deg < \theta < 180^\deg$}                                                 \\
  \MR{2}{KM}          & 5.82/ 4 = 1.46 & 5.47/ 4 = 1.37 & 5.23/ 4 = 1.31 & 5.65/ 4 = 1.41    \\
                      &(5.98/ 4 = 1.50)&(5.58/ 4 = 1.40)&(5.35/ 4 = 1.34)&(5.79/ 4 = 1.45)   \\
  \MR{2}{QGSJET-II-03}& 4.81/ 4 = 1.20 & 4.60/ 4 = 1.15 & 4.48/ 4 = 1.12 & 4.70/ 4 = 1.17    \\
                      &(5.14/ 4 = 1.29)&(4.86/ 4 = 1.22)&(4.72/ 4 = 1.18)&(5.00/ 4 = 1.25)   \\
  \MR{2}{SIBYLL~2.1}  & 8.70/ 4 = 2.17 & 8.88/ 4 = 2.22 & 9.12/ 4 = 2.28 & 8.74/ 4 = 2.19    \\
                      &(9.39/ 4 = 2.35)&(9.49/ 4 = 2.37)&(9.71/ 4 = 2.43)&(9.40/ 4 = 2.35)   \\ \noalign{\smallskip}
\\
  \MC{5}{l}{{\bf ANTARES 2021}  \cite{Albert:2021pwz},
             $10^2 < E_\nu \lesssim 5 \times 10^4$ GeV,
             $90^\deg < \theta < 180^\deg$}                                                \\
  \MR{2}{KM}          & 10.5/ 3 = 3.50 & 12.4/ 3 = 4.14 & 14.0/ 3 = 4.67 & 11.2/ 3 = 3.73  \\ 
                      &(11.2/ 3 = 3.73)&(12.8/ 3 = 4.27)&(14.3/ 3 = 4.75)&(11.8/ 3 = 3.93) \\ 
  \MR{2}{QGSJET-II-03}& 13.8/ 3 = 4.60 & 16.2/ 3 = 5.41 & 18.2/ 3 = 6.05 & 14.8/ 3 = 4.93  \\ 
                      &(13.5/ 3 = 4.50)&(15.6/ 3 = 5.18)&(17.3/ 3 = 5.75)&(14.3/ 3 = 4.77) \\ 
  \MR{2}{SIBYLL~2.1}  & 40.1/ 3 = 13.4 & 43.8/ 3 = 14.6 & 46.6/ 3 = 15.5 & 41.6/ 3 = 13.7  \\ 
                      &(40.0/ 3 = 13.3)&(43.2/ 3 = 14.4)&(45.8/ 3 = 15.3)&(41.3/ 3 = 13.8) \\ \noalign{\smallskip}
 \\
  \MC{5}{l}{{\bf Combined data:}
            {\bf IceCube} \cite{Aartsen:2012uu,Aartsen:2015xup} \&
            {\bf ANTARES 2021} \cite{Albert:2021pwz},
             $10 < E_\nu \lesssim 10^5$ GeV,
             $90^\deg < \theta < 180^\deg$}                                                \\
  \MR{2}{KM}          & 38.8/11 = 3.53 & 40.3/11 = 3.66 & 41.6/11 = 3.78 & 39.4/11 = 3.58  \\ 
                      &(42.1/11 = 3.83)&(43.2/11 = 3.93)& 44.5/11 = 4.05)&(42.5/11 = 3.86) \\ 
  \MR{2}{QGSJET-II-03}& 19.7/10 = 1.97 & 21.8/10 = 2.18 & 23.6/10 = 2.36 & 20.6/10 = 2.06  \\ 
                      &(19.8/10 = 1.98)&(21.5/10 = 2.15)&(23.0/10 = 2.30)&(20.4/10 = 2.04) \\
  \MR{2}{SIBYLL~2.1}  & 49.5/10 = 4.95 & 53.3/10 = 5.33 & 56.3/10 = 5.63 & 51.0/10 = 5.10  \\ 
                      &(50.2/10 = 5.02)&(53.4/10 = 5.34)&(56.2/10 = 5.62)&(51.5/10 = 5.15) \\ \noalign{\smallskip}\hline\hline\noalign{\smallskip}
  \end{tabularx}}
  \end{table*}
The quality of the overall fit can be judged from the global $\chi^2$ divided by ndf. For each data set included in the analysis, a partial $\chi^2/{\rm ndf}$ relating to single experiment is provided. 
The second column in Tables~\ref{Tab:tab-1}--\ref{Tab:tab-3}  presents $\chi^2/{\rm ndf}$ obtained for the  CN flux calculated with hadronic models, KM, QGSJET-II-03 and SIBYLL 2.1 (the flux model indices are dropped).
Columns 3--5 are $\chi^2/{\rm ndf}$ for the total neutrino flux, the conventional and  the prompt one (CN+PN). Here, we show also  results for the three charm production models QGSM, SIBYLL 2.3c, and BEJKRS (see Section II). 

The partial and global $\chi^2/{\rm ndf}$ values for the conventional muon neutrinos illustrate a satisfactory agreement among all the data sets (Tables \ref{Tab:tab-1} and \ref{Tab:tab-2}) except for AMANDA-II data. 
We may state,  that the prompt muon neutrinos predicted with charm production models under study are statistically insignificant (Tables~\ref{Tab:tab-1}, \ref{Tab:tab-2}). More optimistic picture is seen for the contribution of the prompt electron neutrinos with the SIBYLL 2.3c (4th column in Table \ref{Tab:tab-3}): for IceCube-86 experiment $\chi^2/{\rm ndf}$ value  is reduced by $\sim 8-10\%$. Unfortunately, total statistical significance of the $\nu_e$ data is not so high in our analysis due to restricted energy range. 

  \section{Results and discussion}
 
 The obtained $\chi^2$ values for each flux model and each experiment are shown in Tables 
 \ref{Tab:tab-1}--\ref{Tab:tab-3}.  
 Tables~\ref{Tab:tab-1} and \ref{Tab:tab-2} present the values of $\chi^2/{\rm ndf}$ calculated for $\nu_\mu$,  Table~\ref{Tab:tab-3} does the same for $\nu_e$ spectra.
 The energy ranges and cuts for zenith angles are indicated for each experiment.
 Namely, Tables \ref{Tab:tab-1} and \ref{Tab:tab-2} show $\chi^2/{\rm ndf}$ 
 for $\nu_\mu$ spectra by \Frejus\cite{Daum:1994bf}, IceCube-40 2011 \cite{Abbasi:2010ie},
 IceCube-59 2015 \cite{Aartsen:2014qna}, ANTARES 2013 \cite{Adrian-Martinez:2013bqq},
 ANTARES 2021 \cite{Albert:2021pwz},  AMANDA-II 2010 \cite{Abbasi:2010qv},
 and Super-Kamiokande I--IV 2016 \cite{Richard:2015aua}, as compared with the CN and CN+PN
 neutrino spectra  predicted by  QGSJET-II-03 \cite{Kalmykov:1997te,Ostapchenko:2004ss,Ostapchenko:2006wc},   SIBYLL~2.1~\cite{Ahn:2009wx}, and KM \cite{Kalinovsky:1989kk,Kimel:1974sn} for H3a and ZS (in brackets)
 cosmic ray spectra. The CN spectra were averaged over the zenith angles according to the cuts provided by experimentalists. 

The analysis is performed for five combinations: 1) $\nu_\mu$ and $\nu_e$ separately for 
each experiment (${\rm ndf} = 4\div12$);  2) the combined  $\nu_\mu$ data except for IceCube-59 and ANTARES 2021  (${\rm ndf} = 35, 39$) (Table \ref{Tab:tab-1}); 3) combined all $\nu_\mu$ data (${\rm ndf} = 50, 54)$) (Table \ref{Tab:tab-2}); 4) the combined ANTARES 2013 and ANTARES 2021  $\nu_\mu$ data (${\rm ndf} = 15$) (Table \ref{Tab:tab-2}); 5) the combined $\nu_e$ data  of IceCube-79, IceCube-86, and  ANTARES 2021  (Table \ref{Tab:tab-3}) (${\rm ndf} = 10, 11)$).   
 
 In the analysis of Super-Kamiokande data \cite{Richard:2015aua} we use only 4 ($\nu_\mu$) and 2 ($\nu_e$) data points measured at high energies. %
  We consider high energy models  QGSJET-II-03 and SIBYLL~2.1 as reasonable ones 
  at energies $E_\nu \geq 100$ GeV, while KM is valid  in the wider energy range, $E_\nu \geq 10$ GeV.
Thus, we compare the KM-predicted $\nu_\mu$ spectrum ($\text{ndf}=4$) and that of other theoretical models 
($\text{ndf}=2$) with measured ones in \Frejus 1995 \cite{Daum:1994bf} and Super-Kamiokande~\cite{Richard:2015aua} at neutrino energies above 10 GeV. The same relates to four and three points of the $\nu_e$ flux measured in the IceCube~\cite{Aartsen:2012uu}.  %
 
The $\chi^2$ values obtained with KM are QGSJET-II are closely related for all data, differing from those for SIBYLL~2.1.
  QGSJET-II and  KM give the best fit of the IceCube-40 data \cite{Abbasi:2010ie}, 
  while the data obtained by  \Frejus  \cite{Daum:1994bf},  AMANDA-II \cite{Abbasi:2010qv},
  ANTARES 2013 \cite{Adrian-Martinez:2013bqq}, and Super-Kamiokande \cite{Richard:2015aua}
  are better described with SIBYLL~2.1. 
  
The PN fluxes were calculated using all listed charm production models, but only three of them,
  QGSM \cite{Sinegovsky:2018vju},  SIBYLL~2.3c \cite{Fedynitch:2018cbl}, 
  and  BEJKRSS  \cite{Bhattacharya:2016jce}, are presented in Tables~\ref{Tab:tab-1}--\ref{Tab:tab-3}.
As may be seen from these tables, the prompt neutrinos contribution is practically negligible for all measurements.

 Table \ref{Tab:tab-4} presents  central values of the ``crossing energy'' (obtained with six charm production models) which dispersed in the range $0.8-4$ PeV. 
The ``crossing energies'' derived with QGSM  \cite{Sinegovsky:2018vju}, PROSA \cite{Zenaiev:2019ktw} and GM-VFNS \cite{Benzke:2017yjn} are almost coincident for KM (close to that for SIBYLL~2.3c \cite{Fedynitch:2018cbl}) and QGSJET-II-03 \cite{Ostapchenko:2004ss} taken separately. The  GRRST  \cite{Gauld:2015kvh} and  BEJKRSS \cite{Bhattacharya:2016jce} predict highest ``crossing energies''. %
 \begin{table*}[tb]
  \caption{Central values of the crossing energy (in PeV) predicted with the flux models. 
          }
   \label{Tab:tab-4} 
  \center{
  \begin{tabularx}{\linewidth}{lCCCCCK}                                                                                                                                                                                                        \hline\hline\noalign{\smallskip}
  CN $|$ PN models &       QGSM      &   SIBYLL~2.3c   & PROSA & GM-VFNS & GRRST & BEJKRSS  \\ \noalign{\smallskip}\hline      \noalign{\smallskip}
  KM               &       1.70      &      1.59       & 1.65  & 1.59    & 3.97  & 2.98     \\ 
  QGSJET-II-03     &       0.98      &      0.76       & 0.98  & 1.05    & 1.70  & 1.90     \\
  SIBYLL~2.1       &       1.55      &      1.20       & 1.55  & 1.48    & 3.47  & 2.75     \\ \noalign{\smallskip}\hline\hline\noalign{\smallskip}
  \end{tabularx}}
  \end{table*}
  \begin{table*}[t!]
   
  \caption{Comparative significance of CN models
           used in the analysis of $\nu_\mu$ spectra
           derived by IceCube-59 2015 (10 data points), and the combined data of
           ANTARES 2013 \& ANTARES 2021 (15 data points).
           See Table~\ref{Tab:tab-2} for details.
          }
 \label{Tab:tab-5} %
  \center{
  \begin{tabularx}{\linewidth}{lCCK}                                                                                                                                                                                                         \hline\hline\noalign{\smallskip}
  Experiment $|$ CN models                  &  QGSJET-II $/$ KM    &  SIBYLL~2.1 $/$ KM  & SIBYLL~2.1 $/$ QGSJET-II \\ \noalign{\smallskip}\hline \noalign{\smallskip}
  IceCube-59 2015            &   $1.10\, \sigma$      &  $3.63\, \sigma$     &   $3.80\, \sigma $      \\ \noalign{\smallskip}
  ANTARES 2013 + ANTARES 2021 &   $1.07\, \sigma$      &  $3.92\, \sigma$    &   $3.77\, \sigma $      \\ \noalign{\smallskip}\hline\hline\noalign{\smallskip}
  \end{tabularx}}
  \end{table*}

The bulk analysis of conventional muon neutrino spectra shows that all flux models are consistent with measurements. 
However, SIBYLL~2.1 is in a tension with the IceCube-59 data for zenith angles 
 $90^\deg < \theta < 120^\deg$  ($\chi^2/{\rm ndf}\simeq 3.9$), as well as with the ANTARES 2021  
 ($\chi^2/{\rm ndf} \simeq 4.0$) (Table \ref{Tab:tab-2}) -- corresponding $p$ values are
 $p\simeq 6\times10^{-5}$ (IceCube-59) and $p\simeq  10^{-3}$ (ANTARES 2021).
While QGSJET-II and KM  give $p=0.87$ (QGSJET-II) and  $p=0.28$ (KM) for IceCube-59, as well  
  $p=0.999$ (QGSJET-II) and $p=0.88$  (KM) for ANTARES 2021 data. Thus KM and QGSJET-II are in close agreement with the latest measurements of the atmospheric muon neutrino spectrum.  %

 Opposite results were obtained for  AMANDA-II:  SIBYLL~2.1 appears as the preferred model ($p=0.69$) in comparison with KM ($p\simeq 0.01$) and QGSJET-II ($p\simeq 2\times 10^{-4}$).
  
 Notice also that close agreement of the SIBYLL 2.1 prediction ($p=0.999$) with ANTARES 2013 data is ruined by  ANTARES 2021 ($p=1.2\times 10^{-3}$), while KM and QGSJET-II-03 keep the accordance with the latter data ($p\approx 0.88-0.99$). Simialr results are obtained also  for the combined data ANTARES 2013+2021:  $p=0.98$ (KM)  and $p=0.95$ (QGSJET-II) against $p=0.12$ for SIBYLL 2.1  (Tables \ref{Tab:tab-1} and \ref{Tab:tab-2}).
  
The IceCube-59 2015 \cite{Aartsen:2014qna} data within the interval  $90^\deg < \theta < 120^\deg$  are described by models KM, QGSJET-II, and SIBYLL 2.1 with a lower confidence level ($p=0.28$, $0.87$, $6\times 10^{-4}$) as compared with that for angles  $120^\deg < \theta < 180^\deg$ ($p=0.998,\, 0.999$,  and  $0.225$).  The discrepancy may result from an inaccuracy in analysis of events
  induced by neutrinos passing the detector near the horizon. 
  
Our calculations showed that the zenith-angle cut influences moderately on the angle-averaged conventional neutrino flux:   reducing the angle interval by $\sim 1^\deg$ near the horizon leads to decrease in the spectra  by $\sim 3$\% for neutrino energies above 100 TeV. 

Table \ref{Tab:tab-5} presents the comparative statistical significance of hadronic interactions models used in the analysis of $\nu_\mu$ spectra derived in IceCube-59 (10 data points) \cite{Aartsen:2014qna} and in ANTARES experiments \cite{Adrian-Martinez:2013bqq, Albert:2021pwz} (15 points). 
 This table demonstrates the proximity of the QGSJET-II-03 and Kimel \& Mokhov predictions ($\sim 1\sigma$)  relative to these experiments, while  the SIBYLL~2.1 proves a certain tension with the data
  ($ > 3.5\sigma $). 

As regards to $\nu_e$ spectra,  SIBYLL~2.1 and QGSJET-II-03 give a good description of IceCube-79 (with 3 data points involved in the analysis) \cite{Aartsen:2012uu}.  
The KM model gives similar result for the same data (3 points) but fails with $\chi^2/{\rm ndf} \simeq 5.6$ if all 4 points are included. Nevertheless  KM and QGSJET-II-03 give fairly good fit 
 for the IceCube-86 $\nu_e$ spectrum (4 points) \cite{Aartsen:2015xup}: $p\simeq 0.21$ (KM) and  $0.31$ (QGSJET-II).  
 
 Although, for reasons mentioned above, KM seems suitable for describing two of four
  $\nu_e$ data points (beyond 10 GeV) measured with Super-Kamiokande \cite{Richard:2015aua},
  actually the model hardly fits them ($p\simeq 1.5\times 10^{-2}$). 

  The $\chi^2$ analysis of the spectra measured in the IceCube-59 \cite{Aartsen:2014qna},
 IceCube-79 \cite{Aartsen:2012uu}, and IceCube-86 \cite{Aartsen:2015xup}
   shows a slight preference of the  H3a  model for the cosmic ray  spectrum 
 as compared to the ZS parameterization. 

  \section{Conclusions} 
  
   Predicted differential spectra of atmospheric muon neutrino successfully describe 
  the experimental data within the experimental uncertainties. 
Both parameterizations of the cosmic rays spectrum, by Zatsepin \& Sokolskaya and Hillas \& Gaisser, produce close  $\chi^2$ values for the datasets under analysis, i.e. they are statistically undistinguished. 
 
The calculated spectra of atmospheric muon neutrinos agree well with data obtained in \Frejus,
AMANDA, IceCube, and ANTARES experiments.
 QGSJET-II-03 and KM lead to the best description of IceCube-59 data and ANTARES 2021 measurements of the $\nu_\mu$ spectrum, SIBYLL 2.1 is a good  model to describe the AMANDA-II and the ANTARES 2013 muon neutrino data. The Kimel \& Mokhov model also provides suitable predictions  for the IceCube-59 and ANTARES 2021,   and the best one for combined data ANTARES 2013 + ANTARES 2021 ($p=0.98$). The minimal $\chi^2$  value for the total data on the $\nu_\mu$ spectrum is also derived with Kimel \& Mokhov model ($p=0.94$). 

As concerns atmospheric electron neutrinos, low event statistics in the measurements of the $\nu_e$ flux beyond 100 GeV impedes the unique choice of the preferred hadronic interactions model. 
 
The statistical analysis shows that none of the discussed neutrino flux models leads to the  statistically significant conrtibution of the  prompt atmospheric neutrinos in the energy range covered by the neutrino telescopes.

Thus we can infer from the analysis that the high-energy atmospheric neutrino spectra calculated with the consistent scheme  
  \cite{Sinegovskaya:2014pia,Morozova:2017aic,Kochanov:2019owf,
  Morozova:2019hbk, Morozova:2017eeo, Morozova:2017fof, %
  Sinegovsky:2018vju} are sufficiently reliable and  might be suitable for numerical simulation of the atmospheric neutrino events in the  operating neutrino telescopes, as well as in the future experiments, 
Baikal-GVD \cite{Belolaptikov:2021, Avrorin:2021a, Stasielak:2021}, IceCube-Gen2 \cite{Aartsen:2020a, Ishihara:2019a}, and KM3NeT/ORCA \cite{Adrian-Martinez:2016fdl}. 
We expect that increased statistics on the muon, electron and tau neutrino events due to functional capabilities of the next generation of neutrino telescopes will enable one to solve the prompt neutrino problem.    
%

\section*{ACKNOWLEDGMENTS}
We are grateful to D.~Naumov, V.~Naumov, and M.~Sorokovikov for helpful discussions. 
The authors thank K.~Kuzmin for the assistance in computations and censorious remarks. 
 The work of A.~K. is supported by the Russian Federation Ministry of Science
 and Higher Education, project II.16. 
A.~M. was supported by the Russian Scientific Foundation under grant 18-12-00271.
S.~S. acknowledges the support by the Ministry of Science and Higher Education of the Russian Federation under contract FZZE-2020-0017. 


\begin{thebibliography}{99}
  \bibitem{Daum:1994bf}
  K.~Daum \etal \, (\Frejus Collaboration),
 Determination of the atmospheric neutrino spectra
  with the \Frejus detector,
  Z.\ Phys.\ C {\bf 66}, 417 (1995).


  \bibitem{Abbasi:2009nfa}
  R.~Abbasi \etal \, (IceCube Collaboration),
 Determination of the atmospheric neutrino flux and
  searches for new physics with AMANDA-II,
  Phys.\ Rev.\ D {\bf 79}, 102005 (2009)
  [arXiv:0902.0675]. 

  \bibitem{Abbasi:2010qv}
  R.~Abbasi \etal \, (IceCube Collaboration),
 The energy spectrum of atmospheric neutrinos between 2 and 200 TeV
  with the AMANDA-II detector,
   Astropart.\ Phys. {\bf 34}, 48 (2010)
  [arXiv:1004.2357]. 


  \bibitem{Adrian-Martinez:2013bqq}
  S.~Adrian-Martinez \etal \, (ANTARES Collaboration),
  Measurement of the atmospheric $\nu_\mu$ energy spectrum
  from 100 GeV to 200 TeV with the ANTARES telescope,
  Eur.\ Phys.\ J.\ C {\bf 73}, 2606 (2013)
  [arXiv:1308.1599]. 
  
\bibitem{Albert:2021pwz}
  A.~Albert \etal  \, (ANTARES Collaboration),
  Measurement of the atmospheric $\nu_e$ and $\nu_\mu$ energy spectra
  with the ANTARES neutrino telescope,
  Phys. Lett. \ B {\bf 816}, 136228 (2021) 
  [arXiv:2101.12170]. 
    


  \bibitem{Abbasi:2010ie}
  R.~Abbasi \etal \, (IceCube Collaboration),
 Measurement of the atmospheric neutrino energy spectrum
  from 100 GeV to 400 TeV with IceCube,
  Phys.\ Rev.\ D {\bf 83}, 012001 (2011) 
  [arXiv:1010.3980]. 

  \bibitem{Aartsen:2012uu}
  M.~G.~Aartsen \etal \, (IceCube Collaboration),
  Measurement of the atmospheric $\nu_e$ flux in IceCube,
  Phys.\ Rev.\ Lett. {\bf 110}, 151105 (2013) 
  [arXiv:1212.4760]. 
  
  \bibitem{Aartsen:2013eka}
  M.~G.~Aartsen \etal \, (IceCube Collaboration)),
  Search for a diffuse flux of astrophysical muon neutrinos
  with the IceCube 59-string configuration,
  Phys.\ Rev.\ D {\bf 89}, 062007 (2014)
  [arXiv:1311.7048].
  
   \bibitem{Aartsen:2014qna}
  M.~G.~Aartsen \etal \, (IceCube Collaboration),
  Development of a general analysis and unfolding scheme and its application
  to measure the energy spectrum of atmospheric neutrinos with IceCube,
  Eur.\ Phys.\ J.\ C {\bf 75}, 116 (2015) 
  [arXiv:1409.4535]. 

  \bibitem{Aartsen:2015xup}
  M.~G.~Aartsen \etal \, (IceCube Collaboration),
  Measurement of the atmospheric $\nu_e$ spectrum with IceCube,
  Phys.\ Rev.\ D {\bf 91}, 122004 (2015) 
  [arXiv:1504.03753]. 

  \bibitem{Aartsen:2017nbu}
  M.~G.~Aartsen \etal \, (IceCube Collaboration),
  Measurement of the $\nu _{\mu}$ energy spectrum with IceCube-79,
  Eur.\ Phys.\ J.\ C {\bf 77}, 692 (2017)  
  [arXiv:1705.07780].
  

  \bibitem{Richard:2015aua}
  E.~Richard \etal \, (Super-Kamiokande Collaboration),
  Measurements of the atmospheric neutrino flux
  by Super-Kamiokande: energy spectra, geomagnetic effects, and solar modulation,
  Phys.\ Rev.\ D {\bf 94}, 052001 (2016)
  [arXiv:1510.08127]. 
  
  \bibitem{Barr:2004br}
  G.~D.~Barr \etal,
  A three--dimensional calculation of atmospheric neutrinos,
  Phys.\ Rev.\ D {\bf 70}, 023006 (2004)
  [arXiv:astro-ph/0403630].

  \bibitem{Honda:2006qj}
  M.~Honda \etal,
  Calculation of atmospheric neutrino flux
  using the interaction model calibrated with atmospheric muon data,
  Phys.\ Rev.\ D {\bf 75}, 043006 (2007)
  [arXiv:astro-ph/0611418].

  \bibitem{Honda:2011nf}
  M.~Honda \etal,
  Improvement of low energy atmospheric neutrino flux calculation
  using the JAM nuclear interaction model,
  Phys.\ Rev.\ D {\bf 83}, 123001 (2011)
  [arXiv:1102.2688].
 


 \bibitem{Naumov:2000au}
  V.~A.~Naumov, T.~S.~Sinegovskaya,
  Simple method for solving transport equations describing
  the propagation of cosmic ray nucleons in the atmosphere,
  Phys.\ Atom.\ Nucl. {\bf 63}, 1927 (2000).
  
 \bibitem{Naumov:2001} 
 V.~A.~Naumov, T.~S.~Sinegovskaya, Atmospheric proton and neutron spectra at energies above 1-GeV, 
 in \emph{Proceedings of the 27th International Cosmic Ray Conference}  (Hamburg,  2001), Vol. 1, p. 4173 [arXiv:hep-ph/0106015]. 
%

  \bibitem{Kochanov:2008pt}
  A.~A.~Kochanov, T.~S.~Sinegovskaya, and S.~I.~Sinegovsky,
  High-energy cosmic ray fluxes in the Earth atmosphere:
  calculations vs experiments, Astropart.\ Phys. {\bf 30}, 219 (2008)
  [arXiv:0803.2943].
  
   \bibitem{Sinegovskaya:2014pia}
  T.~S.~Sinegovskaya, A.~D.~Morozova, and S.~I.~Sinegovsky,
  High-energy neutrino fluxes and flavor ratio
  in the Earth's atmosphere,
  Phys.\ Rev.\ D {\bf 91}, 063011 (2015)
  [arXiv:1407.3591].
  
  \bibitem{Sinegovsky:2010ijmpa}
S.~I.~Sinegovsky \etal, Atmospheric muon flux at PeV energies, 
Int. J. Mod. Phys. A {\bf 25}, 3733 (2010).

  \bibitem{Kochanov:2013xea}
  A.~A.~Kochanov, T.~S.~Sinegovskaya, and S.~I.~Sinegovsky,
  High-energy cosmic ray muons in the Earth`s atmosphere,
  J.\ Exp.\ Theor.\ Phys. {\bf 116}, 395 (2013)
 
  \bibitem{Kochanov:2019jpcs}
A.~A.~Kochanov \etal, High-energy atmospheric muon flux calculations in comparison with
recent measurements,  J. \ Phys.\ Conf. \ Ser. {\bf 1181}, 012054 (2019)
 [arXiv:1907.00640]. 

\bibitem{izvtas21}
A.~A.~Kochanov \etal, Atmospheric neutrino spectra: a statistical analysis
of calculations in comparison with experiment, 
Bull.\ Russ.\ Acad.\ Sci.\ Phys. {\bf 85}, 433 (2021).

\bibitem{Sinegovsky:2018vju}
  S.~I.~Sinegovsky and M.~N.~Sorokovikov,
  Prompt atmospheric neutrinos in the quark--gluon string model,
  Eur.\ Phys.\ J.\ C {\bf 80}, 34 (2020)
  [arXiv:1812.11341].

\bibitem{Morozova:2017aic}
  A.~D.~Morozova \etal,
  Calculation of atmospheric high-energy neutrino spectra and
  the measurement data of IceCube and ANTARES experiments,
  Bull.\ Russ.\ Acad.\ Sci.\ Phys. {\bf 81}, 516 (2017).

  \bibitem{Kochanov:2019owf}
  A.~A.~Kochanov \etal,
  Examination of calculations of the atmospheric muon and
  neutrino spectra using new measurements,
  Bull.\ Russ.\ Acad.\ Sci.\ Phys. {\bf 83}, 933 (2019).

  \bibitem{Morozova:2019hbk}
  A.~D.~Morozova \etal,
  Influence of cosmic-ray spectrum and hadron--nucleus interaction model
  on the properties of high-energy atmospheric-neutrino fluxes,
  Phys.\ Atom.\ Nucl. {\bf 82}, 491 (2019).

  \bibitem{Morozova:2017eeo}
  A.~D.~Morozova \etal,
  The comparison of the calculated atmospheric neutrino spectra
  with the measurements of IceCube and ANTARES experiments,
  J.\ Phys.\ Conf.\ Ser. {\bf 798}, 012101 (2017).

  \bibitem{Morozova:2017fof}
  A.~D.~Morozova \etal,
 Influence of hadronic interaction models on characteristics
  of the high-energy atmospheric neutrino flux,
  J.\ Phys.\ Conf.\ Ser. {\bf 934}, 012008 (2017).
   

  \bibitem{Kalmykov:1997te}
  N.~N.~Kalmykov, S.~S.~Ostapchenko, and A.~I.~Pavlov,
  Quark-gluon string model and EAS simulation problems at ultra-high energies,
  Nucl.\ Phys.\ B (Proc.\ Suppl.) {\bf 52}, 17 (1997).

  \bibitem{Ostapchenko:2004ss}
  S.~Ostapchenko,
  QGSJET-II: Towards reliable description of very high energy hadronic interactions,
  Nucl.\ Phys.\ B (Proc.\ Suppl.) {\bf 151}, 143 (2006)
  [arXiv:hep-ph/0412332].

  \bibitem{Ostapchenko:2006wc}
  S.~Ostapchenko,
  Hadronic interactions at cosmic ray energies,
  Nucl.\ Phys.\ B (Proc.\ Suppl.) {\bf 175}--{\bf 176}, 73 (2008)
  [arXiv:hep-ph/0612068].


  \bibitem{Ahn:2009wx}
  E.~J.~Ahn \etal,
  Cosmic ray interaction event generator SIBYLL~2.1,
  Phys.\ Rev.\ D {\bf 80}, 094003 (2009)
  [arXiv:0906.4113].


  \bibitem{Kalinovsky:1989kk}
  A.~N.~Kalinovsky, N.~V.~Mokhov, and Y.~P.~Nikitin,
  \emph{Passage of high-energy particles through matter} 
  (American Institute of Physics, New York, 1989).

  \bibitem{Kimel:1974sn}
  L.~R.~Kimel and N.~V.~Mokhov,
  Particle distributions in $10^{-2}$---$10^{12}$ eV energy range
  initiated by high-energy hadrons in dense media,
  Izv.\ Vuz.\ Fiz. {\bf 10}, 17 (1974).



  \bibitem{Zatsepin:2006ci}
  V.~I.~Zatsepin and N.~V.~Sokolskaya,
  Three component model of cosmic ray spectra from 100 GeV up to 100 PeV,
  Astron.\ Astrophys. {\bf 458}, 1 (2006)
  [arXiv:astro-ph/0601475].%

  \bibitem{Gaisser:2012zz}
  T.~K.~Gaisser,
  Spectrum of cosmic-ray nucleons, kaon production, and the atmospheric muon charge ratio,
  Astropart.\ Phys. {\bf 35}, 801 (2012)
  [arXiv:1111.6675].

  \bibitem{Panov:2006kf}
  A.~D.~Panov \etal,
  Elemental energy spectra of cosmic rays
  from the data of the ATIC-2 experiment,
  Bull.\ Russ.\ Acad.\ Sci.\ Phys. {\bf 71}, 494 (2007)
  [arXiv:astro-ph/0612377]

  \bibitem{Panov:2011ak}
  A.~D.~Panov \etal,
  Energy spectra of abundant nuclei of primary cosmic rays
  from the data of ATIC-2 experiment: final results,
  Bull.\ Russ.\ Acad.\ Sci.\ Phys. {\bf 73}, 564 (2009)
  [arXiv:1101.3246].



  \bibitem{Fedynitch:2018cbl}       
  A.~Fedynitch \etal,
  Hadronic interaction model SIBYLL~2.3c and inclusive lepton fluxes,
  Phys.\ Rev.\ D {\bf 100}, 103018 (2019)
  [arXiv:1806.04140].
  
   \bibitem{Zenaiev:2019ktw}                              
  O.~Zenaiev \etal\, (PROSA Collaboration),
  Improved constraints on parton distributions using
  LHCb, ALICE and HERA heavy-flavour measurements and
  implications for the predictions for prompt atmospheric-neutrino fluxes,
  JHEP {\bf 04}, 118 (2020)
  [arXiv:1911.13164]; 
  \url{https://prosa.desy.de/Main_Page}.
  
  \bibitem{Garzelli:2017} 
 M.~V.~Garzelli \etal\, (PROSA Collaboration),
  Prompt neutrino fluxes in the atmosphere with PROSA parton distribution functions,
  JHEP {\bf 05}, 004 (2017)
 [arXiv:1611.03815].
 
 \bibitem{Gauld:2015kvh}         
  R.~Gauld \etal,
  The prompt atmospheric neutrino flux in the light of LHCb,
  JHEP {\bf 02}, 130 (2016)
  [arXiv:1511.06346].


 \bibitem{Bhattacharya:2016jce}                 
  A.~Bhattacharya \etal,
  Prompt atmospheric neutrino fluxes: perturbative QCD models and nuclear effects,
  JHEP {\bf 11}, 167 (2016)
  [arXiv:1607.00193].


  \bibitem{Benzke:2017yjn}                                            
  M.~Benzke \etal,
  Prompt neutrinos from atmospheric charm 
  in the general-mass variable-flavor-number scheme,
  JHEP {\bf 12}, 021 (2017)
  [arXiv:1705.10386].
 

  \bibitem{Kaidalov:1986zs}
  A.~B.~Kaidalov and O.~I.~Piskunova,
  Production of charmed particles in the quark - gluon string model,
  Sov.\ J.\ Nucl.\ Phys. {\bf 43}, 994 (1986).

  \bibitem{Kaidalov:2003au} 
  A.~B.~Kaidalov,
  High-energy hadronic interactions
  (20 years of the quark gluon strings model),
  Phys.\ Atom.\ Nucl.\ {\bf 66}, 1994 (2003).
  
  \bibitem{Bugaev:1989we}
 E.~V.~Bugaev \etal, Prompt leptons in cosmic rays,
  Nuovo Cim.\ C {\bf 12}, (1989).  

\bibitem{Bugaev:1998}
E.~V.~Bugaev \etal, Atmospheric muon flux at sea level, underground and underwater,
 Phys.Rev.\ D {\bf 58},  054001 (1998). 

\bibitem{NSS:1998}
 V.~A.~Naumov, T.~S.~Sinegovskaya, and ~S.~I.~Sinegovsky,
 The $K_{\ell 3}$ form factors and atmospheric neutrino ratio at high energies,
  Nuovo Cim.\ A {\bf 111}, 129 (1998).

 
 \bibitem{Fedynitch:2015} 
  A.~Fedynitch \etal, 
  MCE$_{\rm Q}$ -- numerical code for inclusive lepton flux calculations,
  PoS (ICRC~2015) 1129.
  
 \bibitem{Fedynitch:2016} 
 A.~Fedynitch, Phenomenology of atmospheric neutrinos,
  EPJ Web Conf. {\bf 116}, 11010 (2016).

  

%
%
 
 \bibitem{pval1}
G.~Cowan \etal\, 
 Asymptotic formulae for likelihood-based tests of new physics,
Eur. Phys. J. C {\bf 71}, 1554 (2011). 

\bibitem{pval2} M.~Tanabashi  \etal\, (Particle Data Group), 
Review of Particle Physics, Phys. Rev. D {\bf 98}, 030001 (2018).
 
 \bibitem{Belolaptikov:2021}
 I. Belolaptikov and Zh.-A. Dzhilkibaev on behalf of the Baikal-GVD Collaboration,
 Neutrino telescope in Lake Baikal: Present and nearest future,
  PoS (ICRC2021) 002.
      
 \bibitem{Avrorin:2021a}    
  A.~V.~Avrorin \etal\, 
  High-energy neutrino follow-up at the Baikal--GVD neutrino telescope,
  Astronomy Lett. {\bf 47}, 94 (2021). 
 
 \bibitem{Stasielak:2021}  
J.~Stasielak, P.~Malecki, D.~Naumov \etal\  on behalf of the Baikal-GVD Collaboration, 
High-energy neutrino astronomy -- Baikal-GVD neutrino  telescope in Lake Baikal,
 Symmetry {\bf 13}, 377 (2021). 
 
 \bibitem{Aartsen:2020a}
M.~G.~Aartsen \etal ~(IceCube-Gen2 Collaboration),
IceCube-Gen2: the window to the extreme Universe,
J.\ Phys.\ G: Nucl. Part. Phys. {\bf 48}, 060501 (2021) [arXiv:2008.04323]. 

\bibitem{Ishihara:2019a}   
 A.~Ishihara \etal (IceCube Collaboration),
  The IceCube upgrade - design and science goals, PoS (ICRC2019) 1031 [arXiv:1908.09441].
  \bibitem{Adrian-Martinez:2016fdl}
  S.~Adri\'an-Mart\'inez \etal\, (KM3NeT Collaboration),
  KM3NeT 2.0 -- Letter of intent for ARCA and ORCA, J.\ Phys.\ G: Nucl. Part. Phys. {\bf 43}, 084001 (2016)
 [arXiv:1601.07459]. 
  \end{thebibliography}
 \end{document}